\documentclass[preprint2]{emulateapj}

\newcommand{\HI}{\ion{H}{1}}

\newcommand{\kms}{km\,s$^{-1}$}
\newcommand{\Msun}{M$_\sun$}

\shorttitle{H{\sc i} observations of EA01A/B} \shortauthors{Buyle, De
Rijcke, Dejonghe}

\begin{document}

\title{EA01A/B : high-resolution H{\sc i} imaging of an interacting
pair of post-starburst (E+A) galaxies} \author{Pieter Buyle\altaffilmark{1},
Sven De Rijcke\altaffilmark{1}, Herwig Dejonghe\altaffilmark{1}}

\altaffiltext{1}{Sterrenkundig Observatorium, Universiteit Gent,
Krijgslaan 281, S9, B-9000, Ghent, Belgium; Pieter.Buyle@UGent.be,
Sven.DeRijcke@UGent.be} 

\begin{abstract}
We present high spatial resolution 21~cm HI observations of EA01A and
EA01B, a pair of interacting post-starburst, or E+A, galaxies at $z =
0.0746$. Based on optical HST/WFPC2 images, both galaxies are known to
display disturbed morphologies. They also appear to be linked by a
bridge of stars. Previous HI observations \citep{chang01} had already
uncovered sizable quantities of neutral gas in or near these galaxies
but they lacked the spatial resolution to locate the gas with any
precision within this galactic binary system. We have analysed deep,
high resolution archival VLA observations of the couple. We find
evidence for three gaseous tidal tails; one connected to EA01A and two
emanating from EA01B. These findings confirm, independently from the
optical imaging, that {\em (i)} EA01A and EA01B are actively
interacting, and that, as a consequence, the starbursts that occurred
in these galaxies were most likely triggered by this interaction, and
that {\em (ii)} $6.6\pm 0.9\times10^9$~{\Msun} of neutral gas are
still present in the immediate vicinity of the optical bodies of both
galaxies. The \HI\ column density is lowest at the optical positions
of the galaxies, suggesting that most of the neutral gas that is
visible in our maps is associated with the tidal arms and not with the
galaxies themselves. This might provide an explanation for the
apparent lack of ongoing star formation in these galaxies.
\end{abstract}


\section{Introduction}\label{intro}

Post-starburst galaxies (or PSGs, or E+A galaxies, or k+a/a+k
galaxies) are characterized by their optical spectra, independent of
any morphological or photometric considerations. PSGs contain a
sizable population of very young stars, producing very prominent
Balmer lines, but lack ongoing star formation, hence the absence of
detectable emission lines. This suggests that PSGs are indeed observed
very shortly, within less than 1~Gyr, after the end of a vigorous,
abruptly truncated starburst \citep{dres83,dres99,pog99}. Today, they
constitute only a small fraction of the galaxy cluster population
($<1$\%, \citet{fab91}); at intermediate-redshifts, however, they
constituted a substantial cluster population \citep{bel95}. Although
the trigger and the abrupt end of the starburst is still not fully
understood, photometry of PSGs shows evidence for disturbed
morphologies, e.g. tidal tails, suggesting that in many cases the
trigger is most likely associated to a galaxy-galaxy merger or
interaction \citep{zab96,yang04,blake04,tran03,goto05,yang08}. Using
numerical simulations, \citet{bekki05} has shown that PSGs can be
formed via a major merger of two gas-rich spiral galaxies. In these
merger simulations, a starburst is triggered that consumes the
available gas within a timespan of roughly 1~Gyr, ending the starburst
abruptly. Recent \HI\ observations have uncovered large quantities of
neutral hydrogen gas in a significant fraction of PSGs
\citep{chang01,buyle06,helmboldt07,buyle08}. This opens up the
possibility of investigating the hypothesized merger origin of many
PSGs.


\begin{figure*}
\includegraphics[width=17cm]{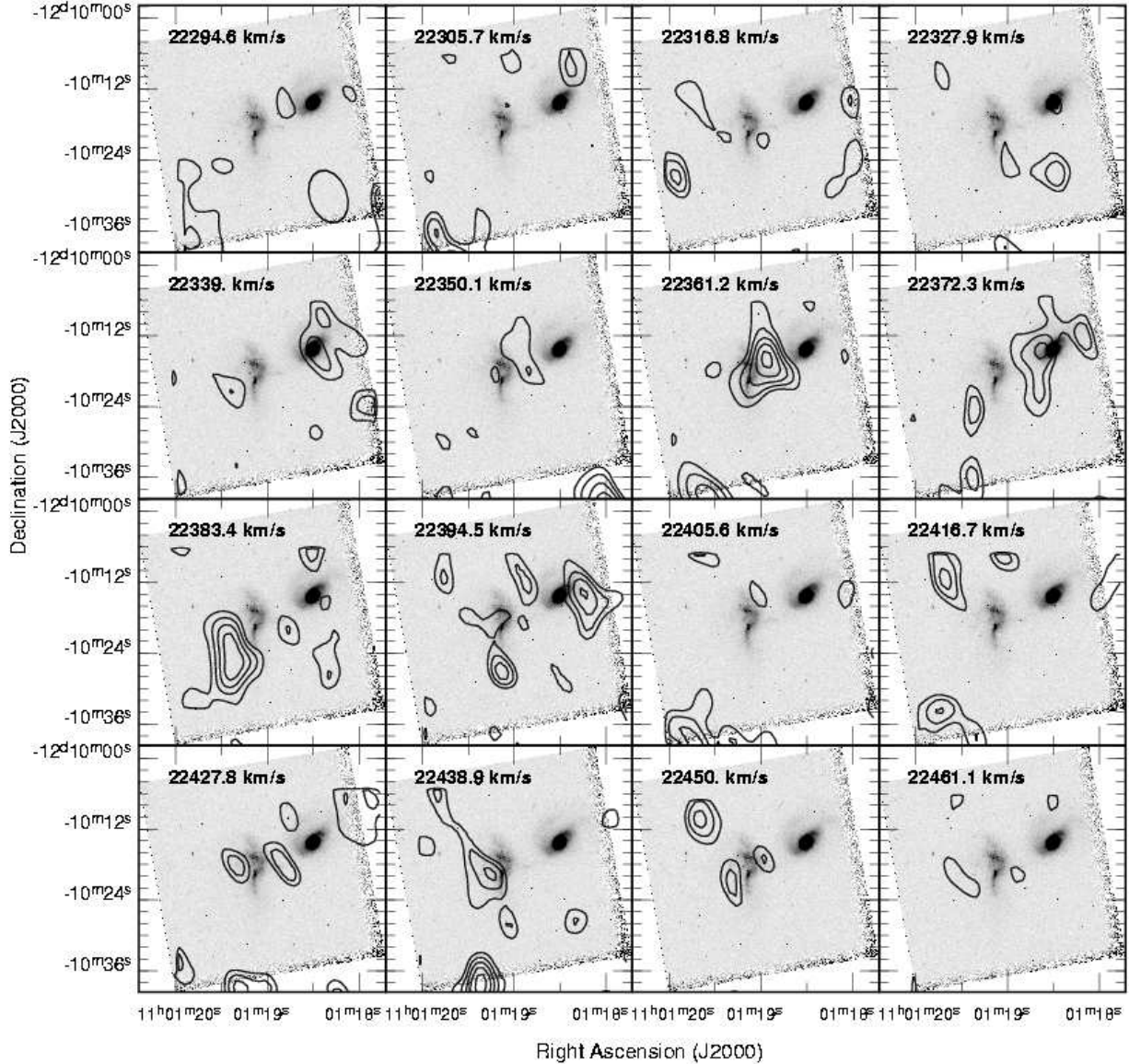}
\caption{Channel maps overplotted on an optical HST/WFPC2 F702W image
of EA01A/B. The contour levels are 2$\sigma$, $3\sigma$,\ldots with
$\sigma$=0.2~mJy~beam$^{-1}$. The synthesized beam is depicted in the
bottom-right corner of the first map (top left).
\label{channels}}
\end{figure*}

Indeed, high-resolution interferometric radio observations of the
neutral gas would provide valuable additional information. On the one
hand, \HI\ morphologies and velocity fields, if they are distorted,
could corroborate the merger hypothesis, and, moreover, could reveal
whether at least part of the neutral gas is still gravitationally
bound and will remain available for future star formation. We selected
the PSG binary EA01A/B as an interesting target for our study. Both
galaxies have a PSG-type optical spectrum and have the same recession
velocity. EA01A is the bluest galaxy of the PSG sample of
\cite{zab96}, suggesting it is also the youngest sample
member. HST/WFPC2 images in the F435W and F702W bands \citep{yang04}
provide strong evidence that the couple is interacting. EA01A,
positioned about 11$''$ east of EA01B, contains many stellar clusters
that are very conspicuous compared with the overall rather low surface
brightness of this galaxy. This diffuse appearance suggests it is on
the verge of disintegrating due to the injected orbital energy. EA01B,
on the other hand, is a bulge dominated early-type spiral galaxy; it
sports strongly asymmetric stellar arms. Both galaxies appear to be
connected by a stellar bridge. Previous HI observations
\citep{chang01} had already uncovered sizable quantities of neutral
gas in or near these galaxies but they lacked the spatial resolution
to locate the gas with any precision within this galactic binary
system.

In section~\ref{obs}, we describe our data reduction and analysis of
deep archival VLA \HI\ observations of the EA01A/B galaxy binary. The
results are presented and discussed in section~\ref{discussion} and
summarized in section~\ref{conclusions}.

\section{Observations and data reduction}\label{obs}

\begin{figure*}
\includegraphics[width=9cm]{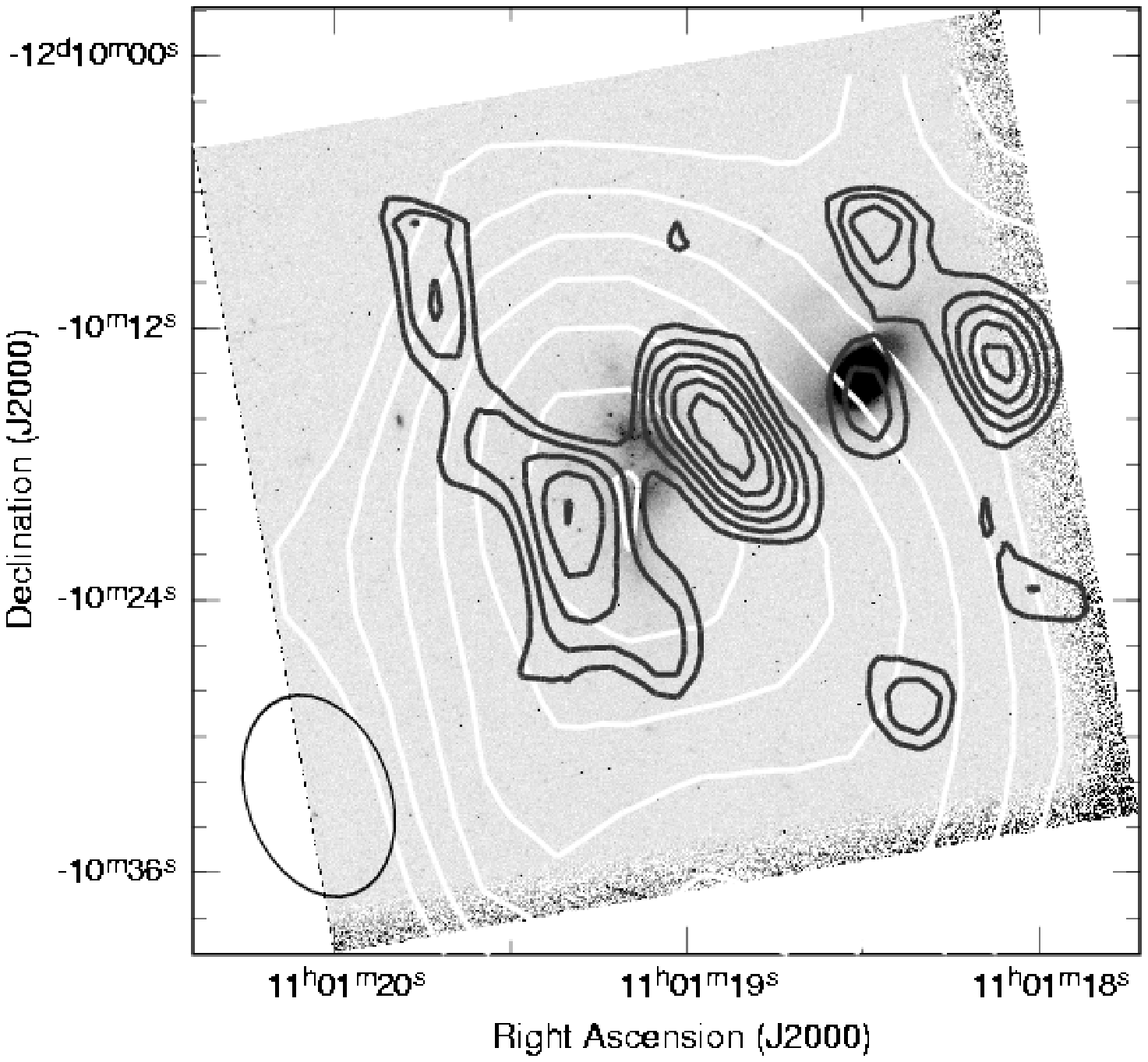}
\includegraphics[width=9cm]{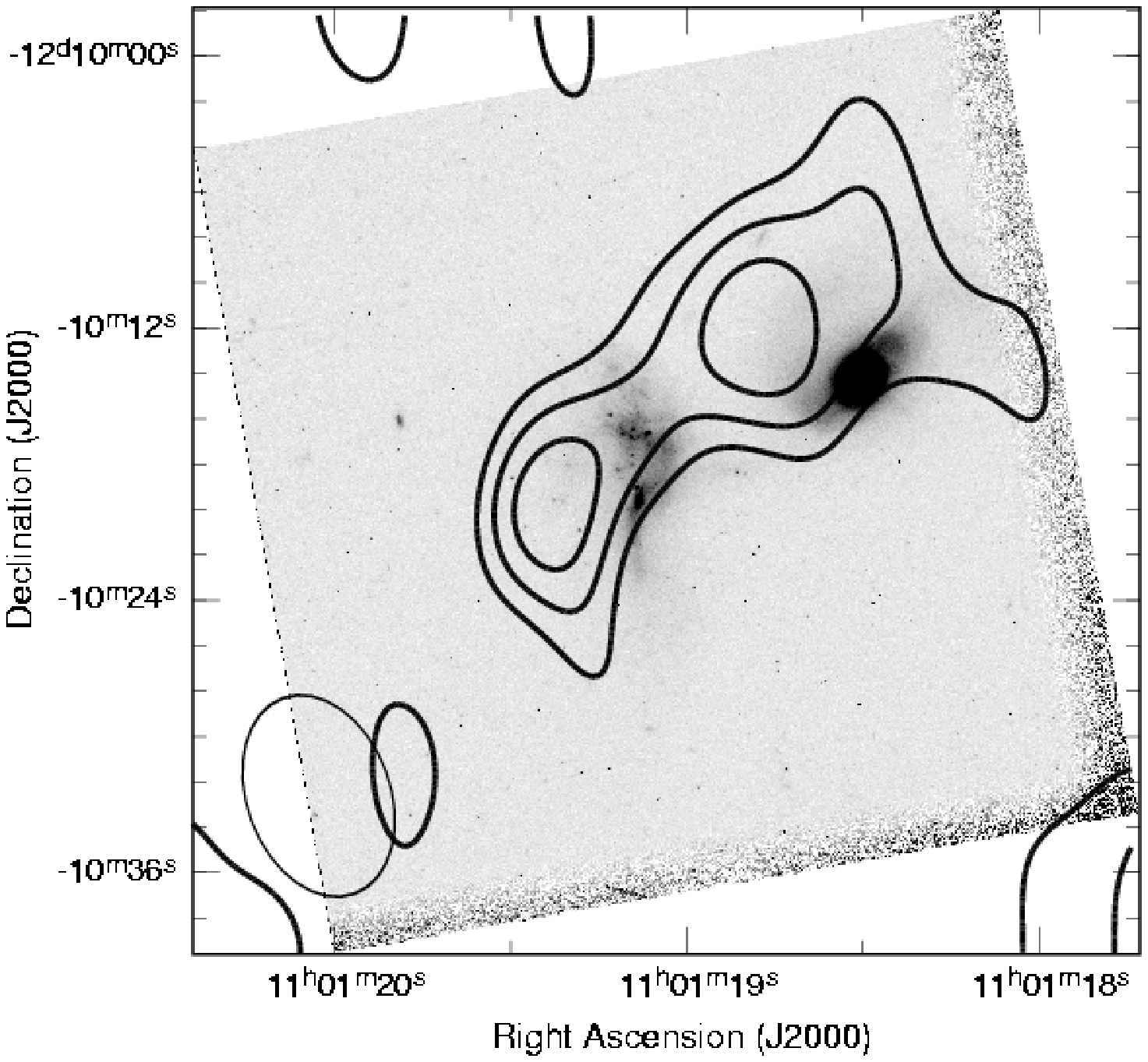}
\includegraphics[width=2.5cm]{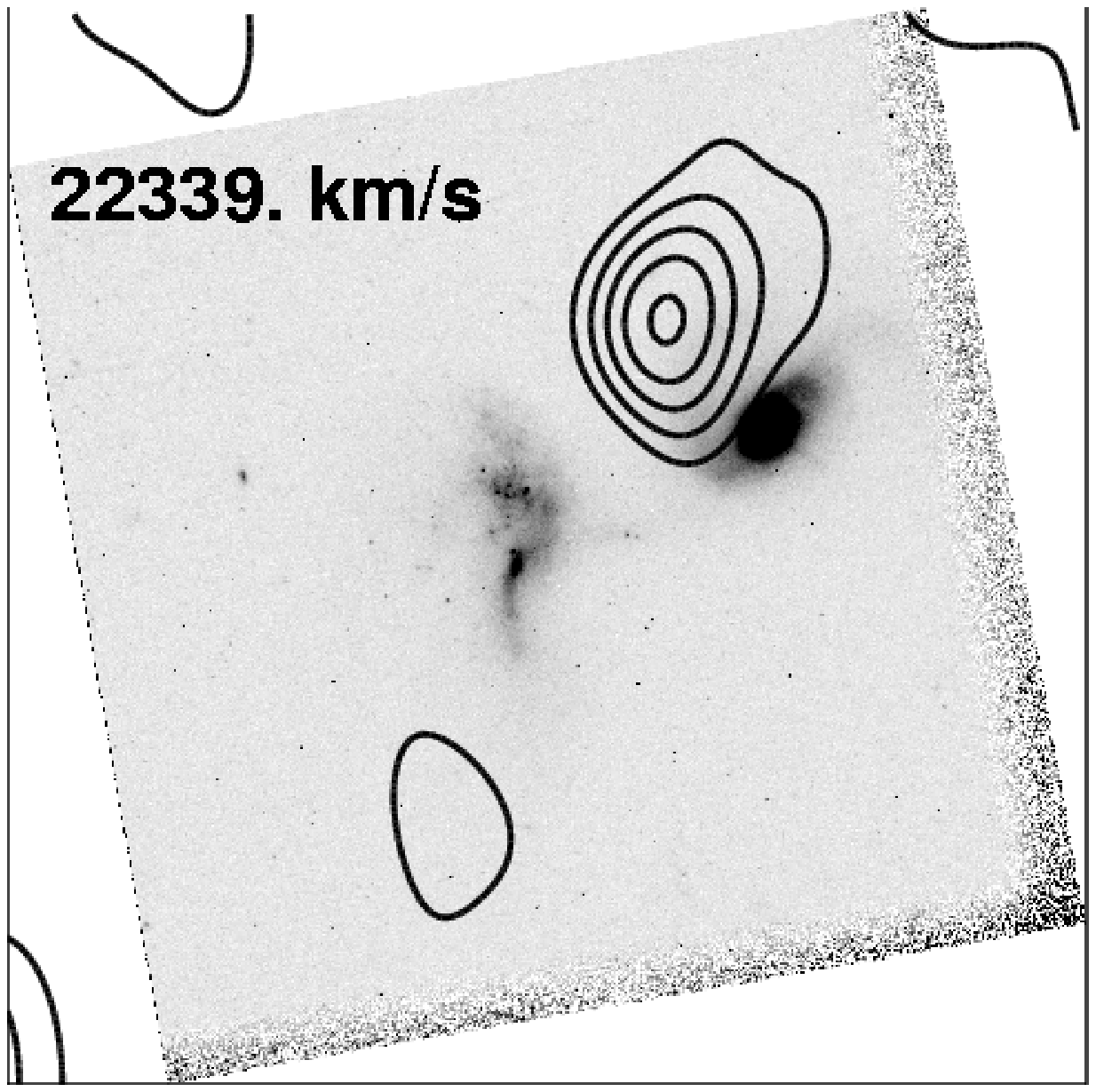}
\includegraphics[width=2.5cm]{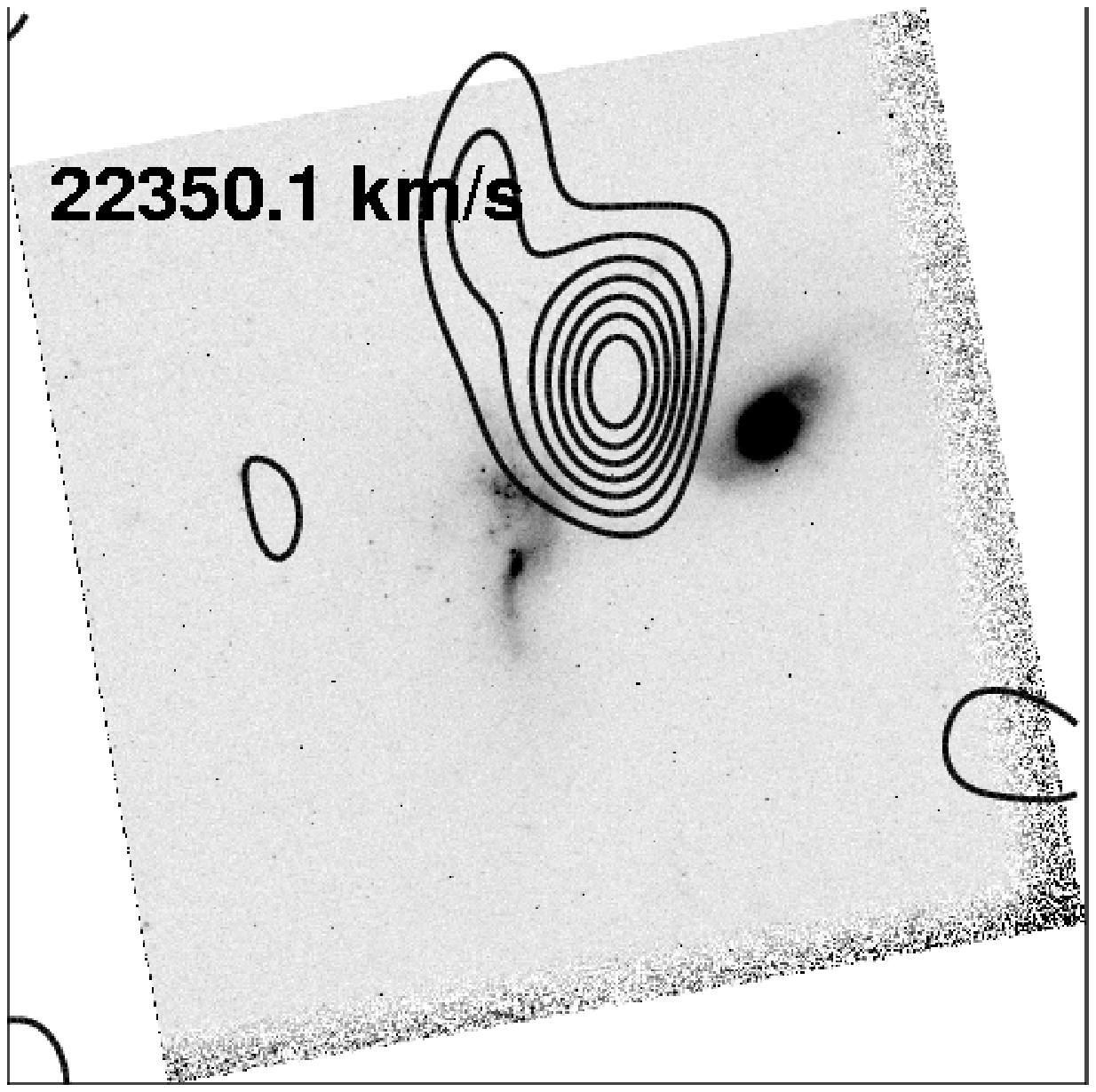}
\includegraphics[width=2.5cm]{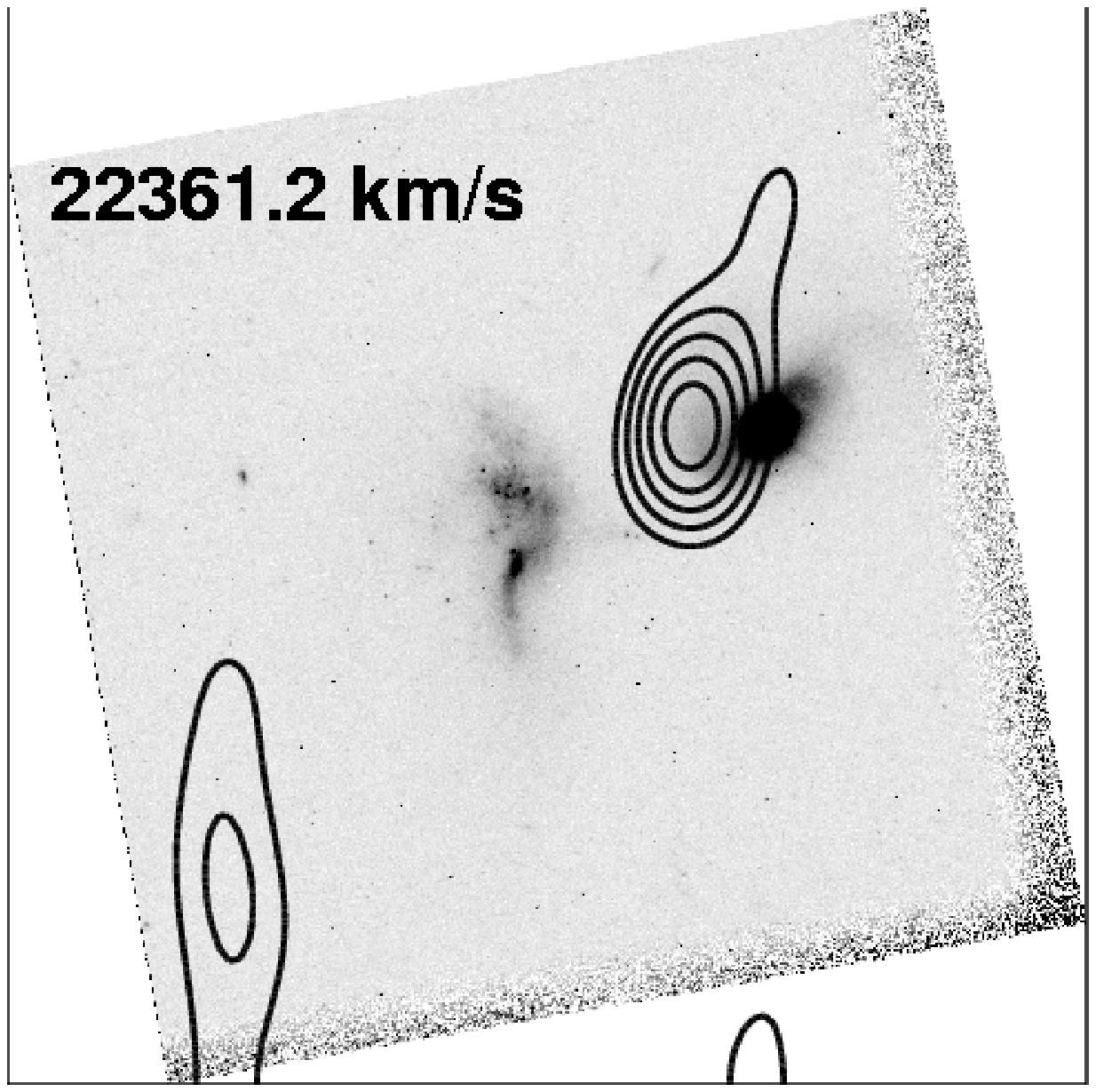}
\includegraphics[width=2.5cm]{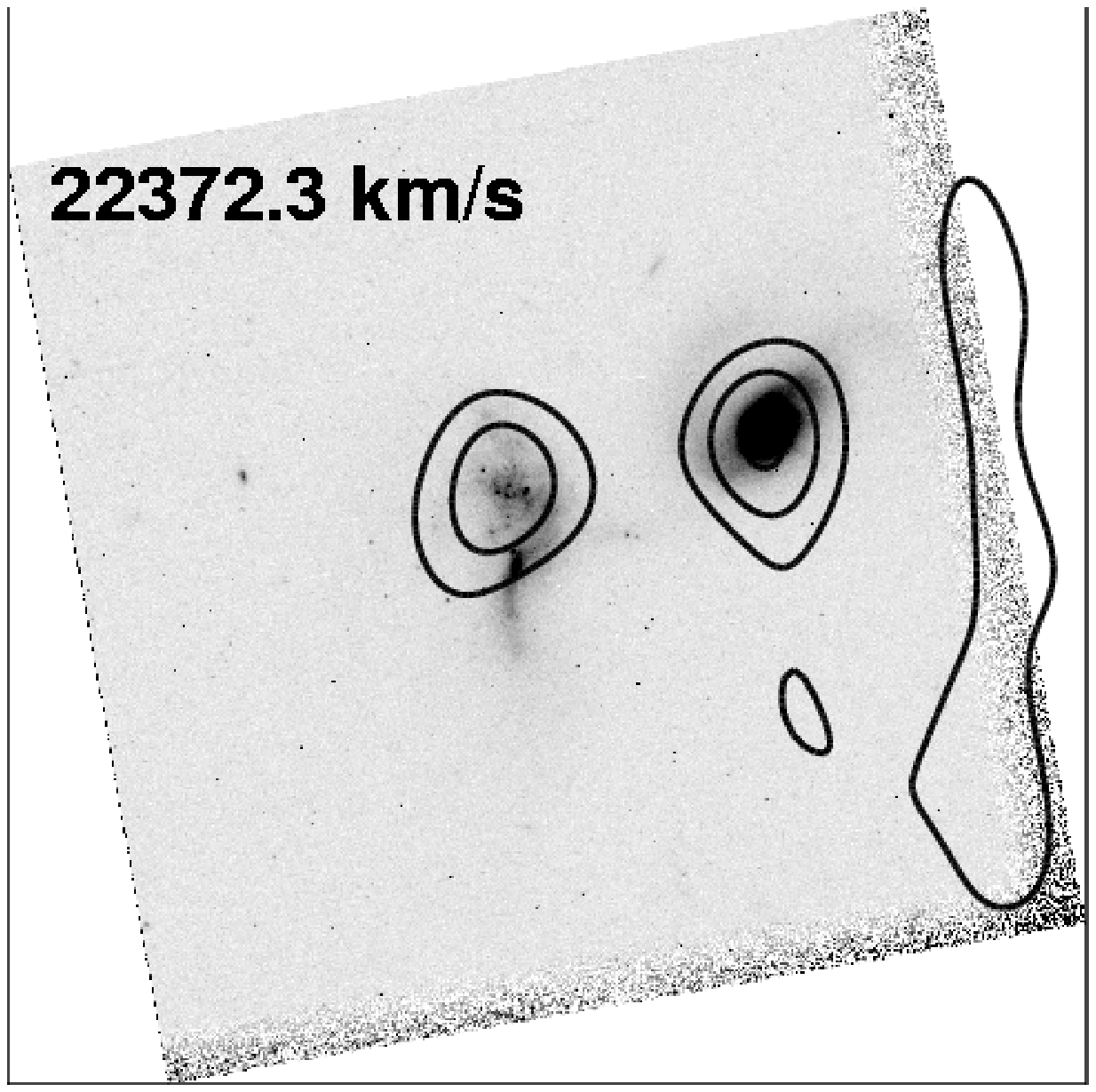}
\includegraphics[width=2.5cm]{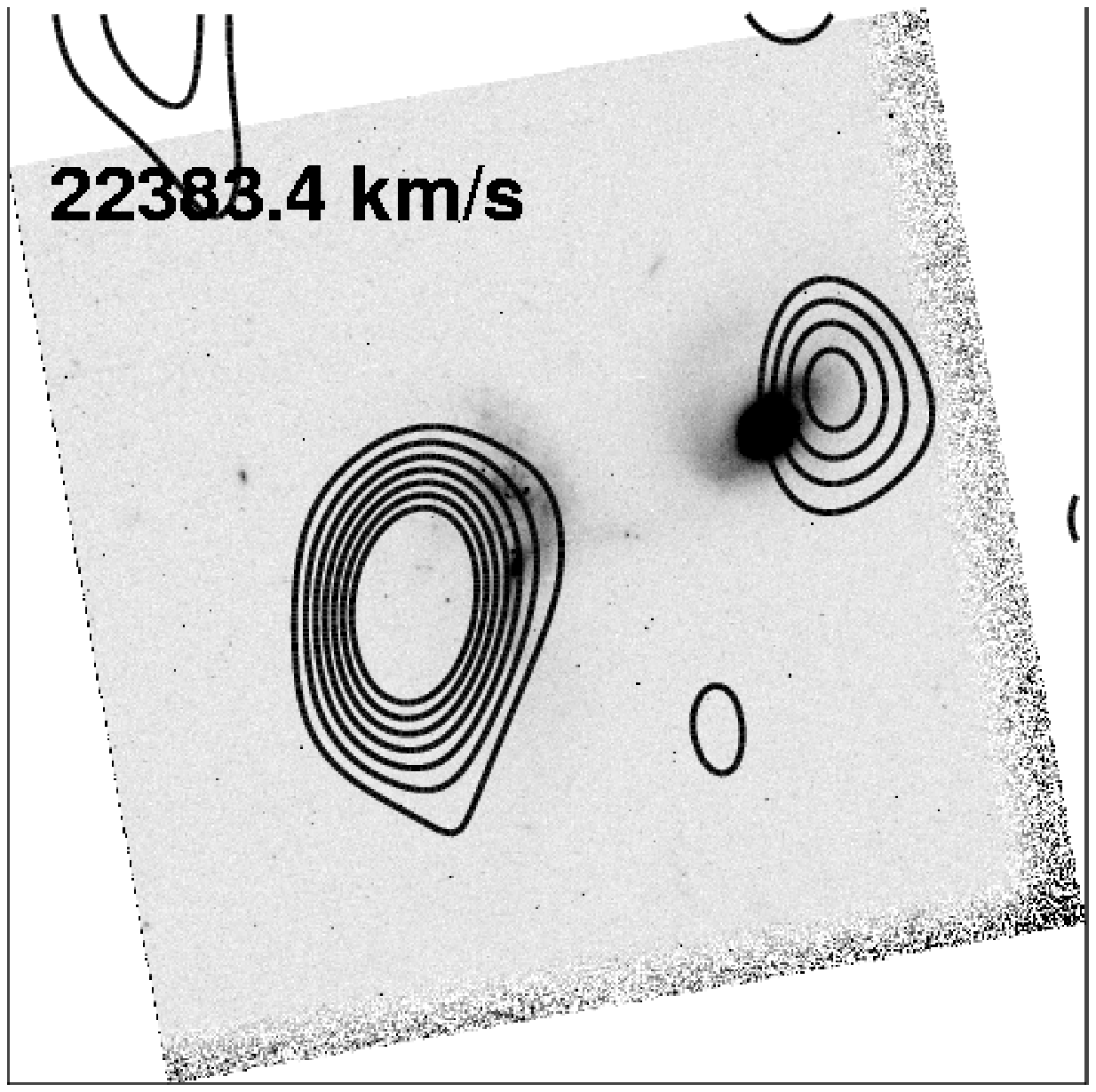}
\includegraphics[width=2.5cm]{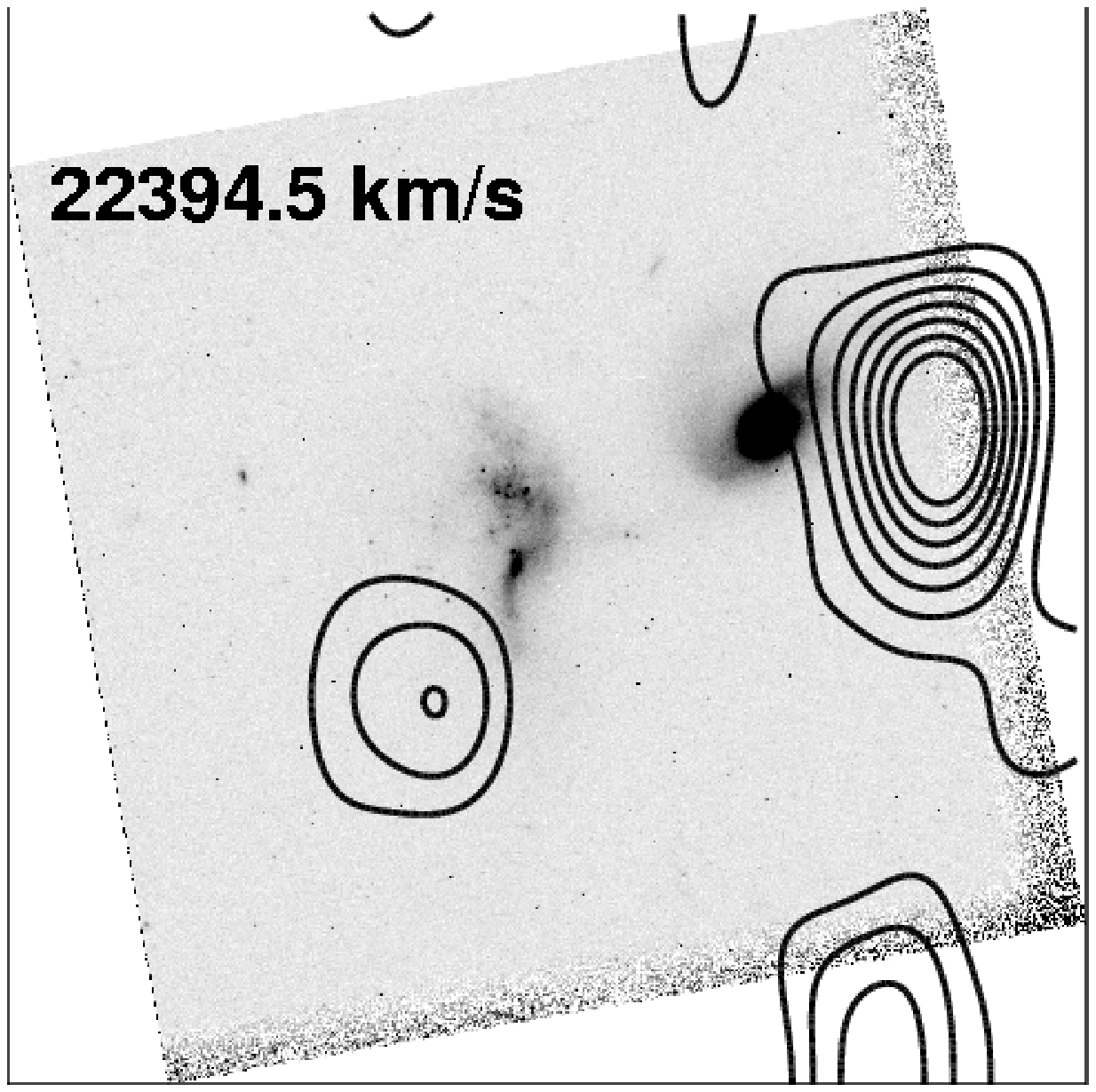}
\includegraphics[width=2.5cm]{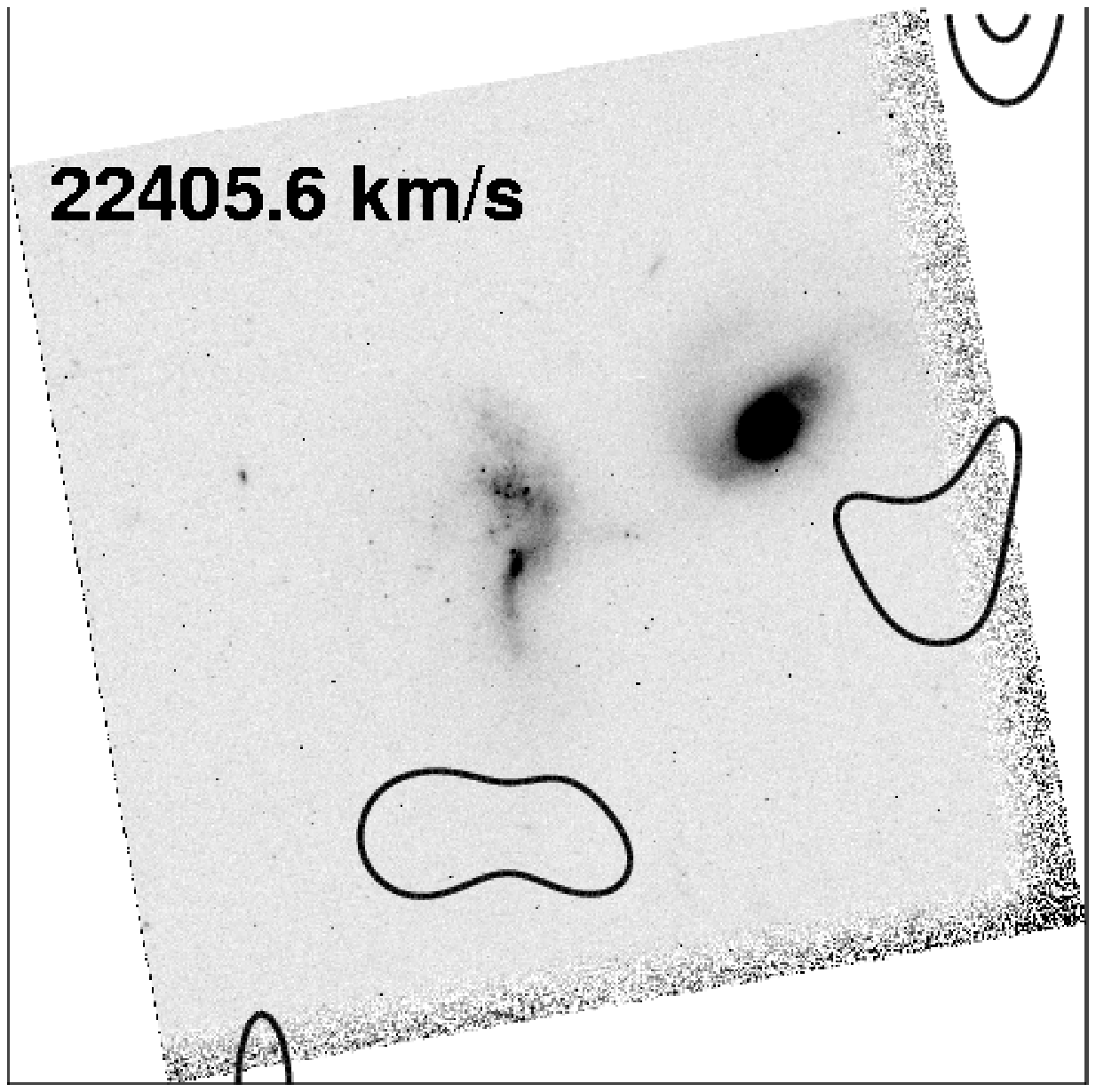}
\caption{The observed \HI\ flux integrated map (left panel) from Chang
  et al. (2001) (white contours) and from this paper (black contours)
  and the \HI\ flux integrated map of a model that roughly reproduces
  the most conspicuous features of the channel maps (right panel). The
  synthesized beam is shown at the bottom left. The contour levels of
  the observed flux map correspond to $2\sigma$, $3\sigma$,\ldots with
  $\sigma=0.01$~Jy~beam$^{-1}$~\kms. The model channel maps are
  displayed in the bottom panels.
\label{mom0}}
\end{figure*}

\begin{table}
\caption{Data characteristics \label{char}}
\begin{tabular}{lc}
\tableline
channel width (km~s$^{-1}$) & 11.1 \\
synthesized beam (FWHM) & 8.87$'' \times$6.32$''$ \\
1$\sigma$ noise level per channel & \\
\hspace{3em} flux density (mJy beam$^{-1}$) & 0.2 \\
\hspace{3em} brightness temperature (K  beam$^{-1}$) & 2.1 \\
\hspace{3em} column density ($\times 10^{19}$ cm$^{-2}$ beam$^{-1}$)  & 4.3 \\
integrated 1$\sigma$ noise level  & \\
\hspace{3em} flux (Jy beam$^{-1}$ km s$^{-1}$) & 0.01 \\
\hspace{3em} brightness temperature (K  beam$^{-1}$ km s$^{-1}$) & 71.1 \\
\hspace{3em} column density ($\times 10^{20}$ cm$^{-2}$ beam$^{-1}$)  & 1.9 \\
\tableline
\tableline
\end{tabular}
\end{table}
We retrieved deep, high-resolution \HI\ 21~cm observations of EA01A/B
from the science archive of the National Radio Astronomy Observatory
(NRAO), made with the Very Large Array (VLA) in New Mexico (USA)
(project number AM0678). The observations were made with the B
configuration during the nights of March 11, 12, 14, 16 and 18,
2001. At a central frequency of $1321.76$~MHz (L-band) and with a
bandwidth of $3.125$~MHz divided over 63 channels, the \HI\ in EA01A/B
was mapped with a velocity resolution of $11.1$~\kms. Alternating
observations of the phase calibrator 1127-145 with a 4 minute exposure
time and EA01A/B with an exposure time of 25 minutes were made,
followed at the end of each night by a 2.5 minute exposure of the flux
calibrator 3C286. This resulted in a total on-source integration time
of $32.6~$h. Standard flagging and calibration of the $u-v$ data was
performed with the Astronomical Image Processing Software (AIPS). The
continuum was subtracted by making a linear fit to the visibilities
over the line-free channels that were not affected by the edge effects
of the band. We created the final datacube using natural weighting to
optimize the sensitivity. This yields a beam-size of $8.87''\times
6.32''$ or $12.3~h^{-1}$~kpc$\times 8.7~h^{-1}$~kpc at the
angular-size distance of EA1 (D$_A=285~h^{-1}$~Mpc, with
$H_0=75~h$~\kms~Mpc$^{-1}$). The rms noise per channel is
$0.2$~mJy~beam$^{-1}$, corresponding to a \HI\ column density of $4.3
\times 10^{19}$ cm$^{-2}$ averaged over the beamwidth. A summary of the
data characteristics can be found in Table \ref{char}.

\section{Discussion}\label{discussion}

We examined the final datacube and detected 21~cm emission in 12
adjacent channels, shown in Fig.~\ref{channels}. The \HI\ flux
integrated map of EA01A/B is presented in the left panel of
Fig. \ref{mom0}. All maps are overplotted onto a HST/WFPC2 F702W image
of the EA01A/B pair. It is clear from these maps that the
\HI\ emission spatially coincides with the optical position of EA01A/B
and is centered around the heliocentric velocity of the previous
\HI\ detection and the optical recession velocity of the pair
($22327.9 -22450.0\, {\rm km}\,{\rm s}^{-1}$).

\begin{figure}
\includegraphics[width=8cm]{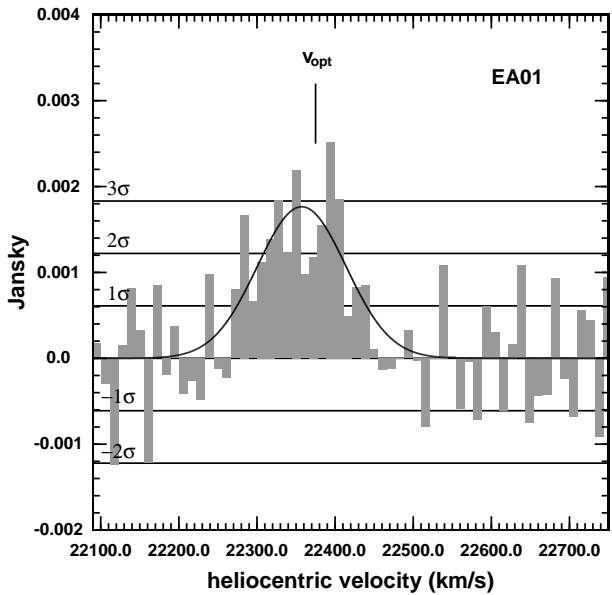}
\caption{The \HI\ profile of EA01A/B, obtained by summing the flux
  within a box around the flux integrated radio emission of the
  galaxy. Rms noise levels are indicated by horizontal lines. A
  Gaussian was fitted to the spectra to measure the velocity widths.
\label{specEA1}}
\end{figure}

The \HI\ profile of EA01A/B, obtained by summing the flux within a box
covering the radio emission of the galaxy is plotted in
Fig. \ref{specEA1}. We fitted a Gaussian to the \HI\ spectrum in order
to derive the total 21~cm line flux. We generated a large library of
mock \HI\ spectra using this Gaussian model and added noise to produce
synthetic data of the same quality as the original spectrum. We then
fitted Gaussian profiles to the mock data in order to estimate the
1$\sigma$ uncertainties on all relevant parameters. We find a total
integrated flux of $S_\nu = 0.26\pm0.03$~Jy~\kms. This is equivalent to an
\HI\ mass of M$_{\rm HI}=6.6\pm 0.9\times10^9$~{\Msun}, adopting a
luminosity distance D$_L$=330~Mpc. The central velocity of the
Gaussian is $22357 \pm 9$~\kms\ and its FWHM is $133 \pm 20$~\kms. We
also re-reduced and analysed the low-resolution \HI\ observations of
\cite{chang01} (Fig. \ref{mom0}). Based on these data, we find a total integrated flux
of $S_\nu=0.29\pm0.08$~Jy~\kms, corresponding to an \HI\ mass of $7.5\pm
2.1\times10^9$~\Msun. The central velocity is $22371 \pm 30$~\kms\ and
the FWHM is $191\pm 88$~\kms. This is in  agreement with the
high-resolution VLA data presented here and with \cite{chang01}.

A plausible interpretation of the \HI\ flux integrated map and the
channel maps is that we are seeing one receding gaseous tidal arm
emanating from EA01A and two tidal arms connected to EA01B, with the
east arm approaching us and the west arm moving away from us. Little
of the detected gas appears to be directly connected with the galaxies
themselves, which show up as minima in the \HI\ column density. This
absence of dense gas may be related to the fact that these galaxies
are in a post-starburst phase. The EA01A arm runs in a south-eastern
direction and can be traced through 3 consecutive channel maps from
22372.3~km~s$^{-1}$ to 22394.5~km~s$^{-1}$. The west EA01B ``arm''
shows up as emission in 3 consecutive channel maps from
22372.3~km~s$^{-1}$ to 22394.5~km~s$^{-1}$. Following its position
through adjacent channel maps with recession velocities below $\approx
22370$~km~s$^{-1}$, the east arm connected with EA01B appears to first
shift in the direction of EA01A and then to return a position slightly
north of EA01B before disappearing in the noise. At that point, the
gas is moving towards us with a radial velocity of about
40~km~s$^{-1}$, relative to the systemic velocity of EA01A/B. The
curvature of the northern arm and the fact that it is approaching us
agrees with the curvature of the stellar arms.

Since the data quality clearly does not allow for a detailed or
quantitative modeling of the gas distribution and kinematics, we
constructed a very simplified model for the \HI\ observations based on
the interpretation outlined above. Two gaseous arms were constructed
to emanate from EA01B, following the position and curvature of the
stellar arms. If the ellipticity $\epsilon = b/a \approx 0.5$ of the
outer isophotes of EA01B is indicative of its inclination, then this
early-type disk is inclined by about 30$^\circ$ from an edge-on
view. We assumed both arms to lie in a disk with the same
inclination. A third gas arm was connected with EA01A, curving in a
south-eastern direction. The \HI\ flux density of these model tidal
arms was assumed to decline exponentially with radius and they were
given solid-body rotation velocity fields in such a way that the mock
channel maps and the flux integrated map derived from this model
provided an acceptable match to the observed maps, taking into account
beam convolution and added noise. Despite its simplicity, this model
can give a fair account of the observations. The model flux map and
channel maps are presented in Fig. \ref{mom0} and can be compared with
Fig. \ref{channels}.

With a 2$\sigma$ density limit of $4 \times 10^{20}$ cm$^{-2}$
beam$^{-1}$, averaged over a $12.3~h^{-1}$~kpc$\times 8.7~h^{-1}$~kpc
beam, we cannot expect to trace the tidal arms very far out. As a
comparison, under these conditions it would be impossible to trace the
southern arm emanating from the interacting pair NGC4038/4039 (``The
Antennae'') further out than $\sim 20$~kpc \citep{hi01}. At the
distance of EA01A/B, this corresponds to $\sim 15''$, less than two
beam diameters. At that distance, the velocity of the NGC4038/4039
tidal arm is less than $\sim 50$~km~s$^{-1}$, in rough agreement with
the largest velocities measured in the EA01A/B system.

\section{Conclusions and summary}\label{conclusions}

EA01A/B is a close pair of post-starburst (E+A) galaxies, surrounded
by some $7 \times 10^9$~{\Msun} of neutral gas. Most of this gas
resides in what appear to be three tidal arms, two of which are
connected with EA01B. Together with optical HST/WFPC2 images, these
observations show that EA01A and EA01B are actively interacting. The
galaxies themselves show up as minima in the \HI\ column density. This
lack of galaxy-bound dense neutral gas is most likely connected with the
fact that these galaxies are in a post-starburst phase.

\acknowledgements We like to thank the referee, Tomotsugu Goto, for
his very useful comments. PB and SDR are post-doctoral fellows of the
Fund for Scientific Research - Flanders, Belgium (F.W.O.). The Very
Large Array is a facility instrument of the National Radio Astronomy
Observatory, which is operated by Associated Universities, Inc. under
contract from the National Science Foundation. This work has made use
of the NASA/IPAC Extragalactic Database (NED) which is operated by the
Jet Propulsion Laboratory, California Institute of Technology, under
contract with the National Aeronautics and Space Administration.

\end{document}